# Effectiveness of Anambra Broadcasting Service (ABS) Radio News on Teaching and Learning (a case study of Awka based Students)

Okechukwu Christopher Onuegbu

*Department of Mass Communication, University of Nigeria, Nsukka, Enugu State, Nigeria.*

**ABSTRACT**
*This work sought to find out the effectiveness of Anambra Broadcasting Service (ABS) Radio news on teaching and learning. The study focused mainly on listeners of ABS radio news broadcast in Awka, the capital of Anambra State, Nigeria. Its objectives were to find out; if Awka based students are exposed to ABS radio; to discover the ABS radio program students favorite; the need gratification that drives students to listen to ABS radio news; the contributions of radio news to students teaching and learning; and effectiveness of ABS radio news on teaching and learning in Awka. The population of Awka students is 198,868. This is also the population of the study. But a sample size of 400 was chosen and administered with questionnaires. The study was hinged on the uses and gratification theory. It adopted a survey research design. The data gathered was analyzed using simple percentages and frequency of tables. The study revealed that news is very effective in teaching and learning. It was concluded that news is the best instructional media to be employed in teaching and learning. Among other things, it was recommended that teachers and students should listen to and make judicious use of news for academic purposes.*

## I. INTRODUCTION

The word 'communication' was derived from the Latin word communicate, which translates to common or shared understanding. This involves both verbal and non-verbal processes of sharing information or ideas between oneself or between one person and another person or group of persons. Thus, it could be said that human communication occurs at various levels or forms such as intrapersonal, interpersonal, group, public, and mass communication.

The most prominent among these forms of communication is mass communication. It is the process of disseminating information or messages to large, anonymous, and scattered heterogeneous audiences via the instrument of mass media, namely, print (newspapers, magazines, journals, books, and other periodicals), and electronics/broadcast (radio, television, and internet).

Asemah and Yaroson (2008), cited in Onuegbu (2016), described mass communication as "The grand and extension of the human communication process; it is human communication on a massive scale, incorporating all the elements of communication-source, encoder, message, channels, decoder, and the receiver, with the seventh element, the feedback. It involves the simultaneous transfer or sending of messages to people in both near and distant places through the use of mass media."

Asemah (2009) enumerates the functions of mass media as a binding influence, entertainment sources, surveillance, persuasion, correlation, the transmission of cultural heritage, selling, and agenda-setting. He further maintains that through the media of mass communication, community members could be bound together, entertained, and induced to accept or buy a particular product or idea.

Radio is a widely used mass communication medium as well as the most effective in dissemination of information, providing education lesson contents, and others to a diverse audience across different geographical locations (Kamat, 2012; United School World, 2020). Experts found that its signals cover almost the entire population. They believed that about 97 percent of the world population is reached by the radio. The radio uses both hardware and software technology for mass production and wide dissemination, which among other things, make it go beyond the immediate proximity of the sender. It inculcates basic knowledge and morality into its audience through interpretation of the messages presented about the environment, prescriptions on what to do about it, and attempts to influence such interpretations, attitudes, and conducts through programs (Onuegbu 2016; Hanxy and Maxcy, 1996).

There is no gainsaying the facts that it performs these functions through its diverse programs like news and current affairs, commentary, documentary, phone-in (which requires the audience to share their opinions/experiences on-air while the program lasts), plays/drama (e.g., Short story), magazines program, commentary, special reports, etc. Thus, Habre & Lockwood (2020) state that "In both urban and rural areas, battery-operated radios broadcast information to entire households. As cheap as $5, radio is less energy-intensive than television and can be shared more easily than a smartphone. The infrastructure was already there. All that educators needed was to adapt content."

*A. Statement of Problem*

The importance of teaching and learning cannot be overemphasized since education is believed to be the bedrock of every society. It is an exercise that could be carried out in formal school settings, at homes, offices, religious and other gatherings. With technological advancement, it could also be conducted online (Internet) or through other mass media like radio, televisions, etc. Radio specifically has various programs like news and current affairs, drama, audience participatory/phone-in-programs, and others. How effective is radio news in teaching and learning?

*B. Objectives of the Study*

The objectives of this study which shall be converted to five research questions, are;
1. This is to find out if Awka based students are exposed to Anambra Broadcasting Service (ABS) radio.
2. To discover the ABS radio program Awka based students favorite.
3. To find the need gratification that drives Awka based students to listen to ABS radio news.
4. To find out the contributions of radio news to students teaching and learning.
5. To find out the effectiveness of ABS radio news on teaching and learning in Awka.

*C. Research Questions*
1. Are Awka based students exposed to the ABS radio?
2. Which ABS radio program do Awka based students expose to?
3. What needs gratification drives Awka based students to listen to ABS radio news?
4. What are the contributions of radio news to students teaching and learning?
5. Is ABS radio news effective in teaching and learning in Awka?

*D. Significance of Study*

This study will inspire educators to amend their academic curriculums in a way that they could easily be included or broadcast in radio news format. It will also enable journalists in the broadcast stations, especially the news writers and newscasters, to cover and report events that are educative.

Similarly, Nigeria and other developing countries could leverage the outcomes of the research to invest more in establishing broadcast stations and in equipping schools with modern facilities so as to easily reach out to the grassroots, especially in emergency situations. This study will also contribute as literature material to scholars and students of mass communication, media studies, and journalism.

**II. A. Theoretical Framework**

This study adopted the uses and gratification theory (UGT). UGT seeks to understand why and how people actively seek out specific media content to satisfy specific needs. It is also seen as an audience-centered approach to understanding mass communication. It was used in this research because researchers, who use this theory, assume that audience members are aware of the impact of media and can articulate their reasons for consuming various media content.

Historically, the Uses and Gratification theory was propounded in the 1940s, when researchers became interested in why people engaged in various forms of media behavior, such as radio listening or newspaper reading. These early studies were primarily descriptive, seeking to classify the responses of audience members into meaningful categories. For instance, Herzog (1944) identified three types of gratification associated with listening to radio soap operas to include emotional release, wishful thinking, and obtaining advice.

In the mass communication process, the uses and gratifications approach puts the function of linking need gratifications and media choice clearly on the side of audience members. It suggests that people's needs influence what media they would choose, how they use certain media, and what gratifications the media give them.

Ideally, several needs and gratification for people are categorized into five. These include Cognitive needs (use of media for acquiring knowledge, news update, information, among other intellectual needs, e.g., quiz programs on radio and television stations), Affective needs (use of media for the satisfaction of all kinds of emotions, pleasure, and other moods), Personal Integrative needs (use of media to realize self-esteem like status, gain credibility and stabilize), Social Integrative needs (to socialize with family, friends, and relations in the society) and Tension free needs (use of the media as a means of escapism and to relieve from tension) (Anunike and Onuegbu, 2020).

*B. Review of Related Empirical Studies*

Empirical researchers have over time focused on the impact of the media (radio news) and its role among students as a learning resource. This section reviews the works of these researchers and their findings.

United World Schools (UWS), a UK registered charity organization, on June 2020, conducted what it called Impact assessment: radio education program (Nepal). The study focused on the radio education program it introduced in ten local radio stations in Nepal in April 2020, probably to sustain teaching and learning in line with the national curriculum amidst lockdown occasioned by the Coronavirus pandemic. In the survey research, over 600 households were interviewed, and the data was analyzed. It showed that most students (72.7% children) were impacted by the program (or found the radio program 'engaging and understood the content').

Al-Mahasneh (2019) investigated the Effect of Video Games on Students' Attention in Government Schools in Jordan. The study, whose purpose was to find the effect of video games on students' attention in government schools, had 150 students in government schools as a sample size. The objectives of the study were to find the effect of video games on students' attention in government schools of Jordan and to find the situation faced by students of government schools of Jordan in this information technology era. The research used quantitative research methodology, and the results were taken by applying mean, regression, and correlation on the data collected from the sample from students studying in the government school. Questionnaires were an instrument for data collection, and each question in the questionnaire has five points or options on the Likert scale. Two variables studied by the researcher were the students' attention which is the dependent variable, and video games which are independent variables. By application of mean, correlation, and regression, the researcher found that both variables are positively related, but the relationship is not significant. The researcher found there was a positive relationship between the two variables and that video games should be allowed for students and advised the educational system to incorporate video games in the educational setting.

BBC Media Action (2017), in a study, How is radio helping to improve girls education in South Sudan, found that listeners of the radio program, Our School – which seeks to raise awareness of the benefits of education and tackle associated barriers, were more involved in education, budgeted for and discussed education with girls or daughters more than non-listeners. The research included a nationwide quantitative survey of 3,169 adults (aged 15 and above) from the former 10 states of South Sudan and several pieces of qualitative research – focus group discussions with audiences, as well as interviews with participants of the community mobilization activities. It conducted a regression analysis to compare key outcomes (such as discussion of education issues) between those exposed and unexposed to Our School while controlling for various factors – confounders – that might have affected the results (such as age or income). It concludes that Our School should continue to focus on addressing increases in ethnic tensions and the impact of the conflict in its content.

Unwin, Weber, Brugha, and Hollow (2017) studied the future of learning and technology in deprived contexts. The research sponsored by Save, the Children International, focuses on the future of basic education, Information Communication Technology (ICT) use in deprived locations, and the use of ICTs in primary school learning in 2020 and 2025, especially in deprived contexts. It draws on research evidence from other researchers (the literature), the authors' experiences of ICT use in education initiatives, interviews with practitioners and academics, a workshop, and consultations with Save the Children staff from many different countries, mostly conducted in August 2017. The research found that there will be considered further divergence in the use of ICTs in classrooms, both between different countries and also within them, in 2020 and 2025 but added that changes in school systems would encourage greater use of technology throughout education. It is recommended, among other things, that all ICT for education programs should begin with a holistic view of education and only then identify the most appropriate technologies to support their delivery.

### III. *A. Research Design*

The research design for this study was a descriptive research type with questionnaires as an instrument for data collection. It is appropriate for this study because it seeks to study the 'effectiveness or impact' of radio news on the audience (teaching and learning). This will make it easier for the researcher to collect and analyze large data within the time allotted to the study.

### *B. Study Area*

Awka, the capital of Anambra State, Nigeria, is the area of study. It is a densely populated area with a population of 301 657 (according to the 2006 national population census). According to an official release obtained from Anambra State Universal Basic Education Board (ASUBEB), Awka has a total of 45 public primary schools and over 65 private or commercial primary schools. Also, information gathered from the Anambra State Post Primary Schools Service Commission (PPSSC) Awka zone shows that Awka has 23 public secondary schools and over 55 private/commercial secondary schools.

Similarly, there are two known tertiary institutions in the capital city, namely, Nnamdi Azikiwe University and Paul's University, and two recognized university study centers; National Open University of Nigeria (Awka Study Center, which is currently at Abagana) and National Teachers Institute (NTI Awka Study Center). Awka also inhibits schools for basic studies, entrepreneurial education institutes, schools for studying foreign languages, vocational studies, skills acquisitions centers, satellite tertiary institutions campuses, among others.

Among the radio stations in the capital city are Anambra Broadcasting Service (ABS radio and television owned by the state government), Federal Radio Corporation of Nigeria (FRCN), and some privately-owned radio stations such as Kpakpando FM, Ogene FM, etc.

### *C. Population of the Study*

The population of this study includes all students based in Awka, notwithstanding their sex group (males and females), races, age groups, educational status, and socioeconomic status. According to recent information obtained from the Anambra State Ministry of Education, the population of Awka students is 198,868.

*D. Sample Size*

A sample size of 400 is selected from 198,868 students populations as claimed by the Anambra State Ministry of Education. The 400 sampling size was arrived at using Taro Yamane Statistical computation developed in 1967 by Taro Yamane, a famous Japanese statistician who contributed greatly in developing sampling methods.

*E. Sampling Technique*

A systematic sampling technique was used in this research. It is a statistical method involving the selection of elements from an ordered sampling frame. Using this procedure, each element in the population has a known and equal probability of selection. To that effect, the researcher visited 10 streets of Awka. At each street, 5 houses were selected, and questionnaire copies were administered to 8 people found listening to the radio at the time of visiting each compound.

*F. Method of Data Collections*

The method of data collections applied in this research was quantitative, which include administering closed-ended questionnaires on respondents or students residing in Awka capital territory across the following axis in the town, namely; Kwata/Unizik Temporary site junction, Eke-Awka Market, Amawbia bypass, Arroma junction, Ifite-Awka/Unizik perm site, and Nkwo Amenyi market. The decision was taken because the areas mentioned witness a large number of students on a daily basis.

*G. Instrument for Data Collection*

The questionnaire is the instrument for data collection in this research. There are two types of questionnaires, namely, structured or closed-form and unstructured or open-ended. But the researcher adopted a structured or closed form.

*H. Validation of Instrument*

In order to ensure that the information collected from a study is true, reliable, and unbiased, the Student Engagement Instrument (SEI) is used in validating this instrument. SEI is a student self-report survey designed to measure cognitive and affective engagement based on a model of engagement that grew out of work with Check & Connect.

Check & Connect mentors recognized that successfully re-engaging students required more than meeting the academic and behavioral standards of schools (Christenson, 2008). These include attention to students' cognitive (e.g., self-regulation, perceived relevance of schooling, future goals) and affective engagement (e.g., belonging, relationships with teachers and peers) at school and with learning.

*I. Reliability of Instrument*

To test the reliability of this study, the researcher conducted a pilot study using 20 respondents randomly selected from the population. The instrument was administered to the respondents in Nnewi, Nnewi LGA of Anambra State, who fill and return them. The analysis helped the researcher establish the validity of the instrument.

To further confirm its reliability (consistency), the researcher re-administered the same instrument with another 20 respondents at Onitsha after two weeks of the first exercise. Analysis of their answers this time showed that there was no difference between the answers they supplied in the first exercise and those of the second. This convinced the researcher of the instrument's reliability.

**IV. A. Method of Data Analysis**

The method of data analysis used in this research was simple percentages and frequency tables. This was adequately adapted to erase the difficulties in understanding the results obtained through the investigation. A total of four hundred (400) questionnaires were produced and administered to 400 respondents. But 351 were duly answered and returned to the researcher while forty-nine (49) were missing. The results were analyzed in the tables below.

**Table 1: Gender Distribution of Awka Students**

| *Sex* | *No. of Respondents* | *Percentages (%)* |
|---|---|---|
| Male | 144 | 41% |
| Female | 207 | 59% |
| TOTAL | 352 | 100% |

Table 1 shows that out of 351 respondents, 144 representing 41% are males, while 207 (59%) are females.

**Table 2: The Respondents Educational Qualifications**

| *Educational Qualification* | *No. of Respondents* | *Percentages (%)* |
|---|---|---|
| FSLC/SSCE | 89 | 25% |
| OND/NCE | 98 | 28% |
| HND/B.Sc./B.A/B. Tech. | 85 | 24% |
| PGD/PhD | 79 | 23% |
| TOTAL | 351 | 100 |

As indicates in table 2, out of three hundred and fifty-one (351) respondents who attempted the questionnaire, 89 representing 25%, hold First School Leaving Certificate (FSLC) and Senior Secondary School Certificate (SSCE), 98 or 28% have acquired Ordinary National Diploma (OND) and National Certificate on Education (NCE) respectively. Also, 85 respondents representing 24% have Higher National Diploma (HND) and first Graduate Degrees (B. Sc./BA/B. Tech), even as holders of Masters of Art (MA), Masters of Science (MSC), and Doctorate Degree (Ph.D.) are 79 (23%).

**Table 3: Respondents who are teachers and students**

| *Are you a student or teacher?* | *No. of Respondents* | *Percentages(%)* |
|---|---|---|
| Yes | 282 | 80% |
| No | 69 | 20% |
| TOTAL | 351 | 100% |

In table 3, out of 351 respondents, 282 or 80% say they are students/teachers, while 69 (20%) said 'No'. As a result, 282 responses were used in presenting and analyzing table 4 below because their answers show that they constitute the population of the study.

**Table 4: Respondents place of study or teaching**

| Schools | No. of Respondents | Percentages (%) |
|---|---|---|
| Secondary School | 114 | 40% |
| Vocational/Basic studies | 41 | 15% |
| Tertiary institutions | 127 | 45% |
| TOTAL | 282 | 100% |

In Table 4 above, out of 282 respondents, a total of 114 respondents representing 40% teach/study in secondary schools, 41 respondents (15%) are in Vocational/Basic Studies, while 127 respondents representing 45% study/teach in tertiary institutions such as universities, colleges of education and polytechnics.

**Table 5: Respondents visiting or residing in Awka**

| Respondents addresses | No. of Respondents | Percentages (%) |
|---|---|---|
| Yes | 282 | 100% |
| No | 0 | 0% |
| Total | 282 | 100% |

According to table 5 above, the whole 282 respondents (100%) who attempted the question admit that they are visitors and residents of Awka, and zero respondents (0%) answered no.

**Table 6: Respondents who listen to the radio**

| Do you listen to Radio? | No. of Respondents | Percentages (%) |
|---|---|---|
| Yes | 270 | 96% |
| No | 12 | 4% |
| TOTAL | 282 | 100% |

As states in table 6, out of 282 respondents, 270 (96%) listen to the radio while 12 (%) do not. Therefore, the 270 respondents that answered yes would be used in analyzing table 7 below.

**Table 7: Respondents favorite radio stations**

| Radio stations | No. of Respondents | Percentages (%) |
|---|---|---|
| ABS radio | 73 | 27% |
| Purity Fm | 76 | 28% |
| Unizik FM | 51 | 19% |
| All of the above | 70 | 26% |
| TOTAL | 270 | 100% |

Table 7 shows that out of 270 respondents that listen to the radio, 73 or 27% are exposed to ABS radio, 76 (28%) tune in to Purity FM, 51 (19%) are for Unizik FM, while 70 (26%) listen to all the radio stations at their convenient time. As a result, the 73 respondents who listen to ABS and 70 that listen to all the radio stations were added to make up 143 respondents used in analyzing table 8 below. This is because they have knowledge of the subject matter.

**Table 8: Respondents favorite ABS radio program**

| ABS radio programs | No. of Respondents | Percentages (%) |
|---|---|---|
| News/current affairs | 40 | 28% |
| Sports Commentaries | 42 | 29% |
| Phone-In/Discussion program | 27 | 19% |
| Drama | 34 | 24% |
| TOTAL | 143 | 100% |

As presents in table 8, out of 143 respondents who listen to ABS, 40 representing 28%, said that their favorite program is News/Current Affairs, 42 (29%) retort that Sports commentaries were their choice program, 27 (19%) are for Phone-In/Discussion program, while 34 (24%) are for Drama program. Thus, the 40 respondents that chose ABS news were used in analyzing subsequent tables because they have knowledge of the questions therein.

**Table 9: Respondents' frequency of listening to ABS Radio News Program**

| Frequency | No. of Respondents | Percentages (%) |
|---|---|---|
| Often | 10 | 25% |
| Very often | 25 | 62% |
| Undecided | 5 | 13% |
| TOTAL | 40 | 100% |

In table 9, out of 40 respondents who are regular listeners to ABS radio news programs, 10, which accounts for 25%, say they do that often, 25 (62%) very often, while 5 (13%) did not decide on what to say.

**Table 10: Media functions of ABS news**

| ABS news functions | No. of Respondents | Percentages (%) |
|---|---|---|
| Information | 10 | 25% |
| Education | 10 | 25% |
| Entertainment | 5 | 13% |
| All of the above | 15 | 37% |
| TOTAL | 40 | 100% |

As contains in table 10, out of 40 respondents that listen to ABS news, 10 representing 25% say it informs, 10 (25%) say it educates, 5 (13%) argue that it entertains, while 15 (37%) said that news does all the media functions.

**Table 11: Respondents assessment of ABS News**

| ABS news assessments | No. of Respondents | Percentages (%) |
|---|---|---|
| Impactful | 40 | 100% |
| No impact | 0 | 0% |
| TOTAL | 363 | 100% |

In table 11, the whole 40 respondents (100%) that attempted the questions admitted that ABS radio news is impactful, while zero (0%) said no.

**Table 12: ABS news contribution to teaching and learning**

| ABS news functions | No. of Respondents | Percentages (%) |
|---|---|---|
| Positively | 29 | 73% |
| Negatively | 4 | 10% |
| Undecided | 7 | 17% |
| TOTAL | 40 | 100% |

As contains in table 12, out of 40 respondents that accept that ABS news contributes to teaching and learning, 29 (73%) said it contributes positively, 4 (10%) admit that it contributes negatively, while 7 (17%) were undecided.

**Table 13: Professions/fields ABS news contribute most**

| ABS news contributions | No. of Respondents | Percentages (%) |
|---|---|---|
| Sciences | 3 | 7% |
| Social Sciences | 5 | 13% |
| Law | 2 | 5% |
| Arts/Humanities | 7 | 17% |
| Agriculture | 4 | 10% |
| Languages | 10 | 25% |
| Education | 8 | 20% |
| Craft/other skills | 1 | 3% |
| TOTAL | 40 | 100% |

Out of 40 respondents who talked about the contributions of ABS radio news to teaching and learning in Table 13 above, 3 representing 7% say it contributes most on Sciences, 5 or 13% stand for Social Sciences, 2 (5%) for Law, 7 (17%) opined that it is good for Arts/Humanities, 4 or 10% said it is good for Agriculture, 10 (25%) were for Languages, Education account for 8 (20%), and 1 (3%) maintain that it contributes to teaching and learning Craft/other skills.

**Table 14: Effectiveness of ABS news to teaching and learning**

| Effectiveness of news | No. of Respondents | Percentages (%) |
|---|---|---|
| Effective | 14 | 35% |
| Not Effective | 6 | 15% |
| Very Effective | 20 | 50% |
| TOTAL | 40 | 100% |

In Table 14 above, out of 40 respondents that attempted the question, 14 (35%) affirm that ABS news contributes effectively to teaching and learning, 6 (15%) said it is not effective, while 20 (50%) said it is very effective.

*B. Discussions of Findings*

The findings of this study were discussed using research questions;

*Research questions*
*Question one*:
What are the levels of exposure of Awka based students to Anambra Broadcasting Service radio?
Data in table 6 provides an answer to this research question. It reveals that the majority of Awka based students listen to radio stations as a total of 270 (96%) respondents out of 282 admit that they exposed themselves to Anambra State Broadcasting Service, while 12 (4%) do not. Also, in table 7, 73 or 27% of respondents say they are exposed to ABS radio, 76 (28%) tune in to Purity FM, 51 (19%) are for Unizik FM, while 70 (26%) listen to all the radio stations at their convenient time. This supports Familusi and Owoeye (2014) research on 'An Assessment of the Use of Radio and other Means of Information Dissemination by the Residents of Ado- Ekiti, Ekiti-State, Nigeria' whereby 118(98%) respondents said they are much exposed to radio programs.

*Question Two*:
What are the Awka based student's Anambra Broadcasting Service favorite programs?
Data in Table 8 supplies answer to this question. It reveals that 40 representing 28% respondents are exposed to ABS News/Current Affairs, 42 (29%) favorite Sports commentaries, 27 (19%) are for Phone-In/Discussion programs, while 34 (24%) are for Drama programs. In other words, lovers of the ABS news program were second. This finding was in support of Familusi, and Owoeye (2014) research on 'An Assessment of the Use of Radio and other Means of Information Dissemination by the Residents of Ado- Ekiti, Ekiti-State, Nigeria' whereby70 respondents (58%) agree that they tune-in to radio mostly for news, while 74 (62%) respondents strongly agree.

*Question Three:*
What needs gratification that drives Awka based students to listen to Anambra Broadcasting Service radio news?
Data in table 10 provides an answer to this question. In the table, 10 respondents representing 25% say they derive information from the ABS news, 10 (25%) say it educates, 5 (13%) argue that it entertains, while 15 (37%) said that news does all the media functions. It re-echoed the students' need gratification for using social media as contained in a study entitled 'Social Media Use among Students of Universities in South-East Nigeria' conducted by Ezeah, Asogwa, and Edogor (2013). The research which used survey research method found that students need gratification for using social media were watching movies (films) and pornographies, for discussion of serious national issues like politics, economy and religious matters, and means of communicating to their leaders on national affairs.

*Question four:*
What are the contributions of radio news to students teaching and learning?

Data in table 12 supplies the answer to this research question. In the table, 29 (73%) respondents say that ABS news contributes positively to teaching and learning, 4 (10%) argue that it does not (that is, it contributes negatively), while 7 (17%) were undecided. Furthermore, table 13 reveals that ABS radio news contributes mostly on the following, Sciences (3 or 7% respondents affirm to that), Social Sciences (5 or 13% respondents agreed), Law (2 or 5% respondents), Arts/Humanities (7 or 17% respondents attested to that), Agriculture (4 or 10% respondents admit), Languages (10 or 25% respondents admit), Education (8 or 20% respondents account to that), and Craft/other skills (1 or 3% accepted). This was in support of Chandar and Sharma (2003) survey research on Bridges to Effective Learning Through Radio conducted in Indira Gandhi National Open University, India, where 50 percent of respondents, when asked about the role radio plays in their daily life, indicated they received information that influences their lives, whilst 25 percent indicated they primarily used radio for learning purposes.

*Question five:*
How effective is the Anambra Broadcasting Service radio news on Awka based students?
Data contained in Table 14 answered this research question. It revealed that 14 (35%) respondents agree that ABS news contributes effectively to teaching and learning, 6 (15%) said it is not effective, while 20 (50%) said it is very effective. It agrees with Djankov, McLiesh, Nenova, and Shleifer's (2001) survey research on media ownership in 97 countries around the world. In the report, 68% of respondents said they listen to radio news often.

*C. Summary of findings*
The following findings were made in this study:
1. That majority of Awka based students tune-in or listen to Anambra Broadcasting Service (ABS) radio broadcast every day;
2. That majority of Awka based students favorite sports programs of ABS radio station;
3. That the need gratification that drives Awka based students to tune in or listen to Anambra Broadcasting Service radio news is education;
4. That the contributions of radio news to students teaching and learning are positive, and mostly in the field of Languages.
5. That ABS news is effective on Awka based students teaching and learning.

*D. Conclusion*
The study examined the effectiveness of Anambra Broadcasting Service (ABS) radio news on teaching and learning (a case study of Awka based students). It concludes that the ABS radio news is effective in teaching and learning.

*V. A. Recommendations*
Having discovered that the news broadcast of Anambra Broadcasting Service is effective in teaching and learning, the researcher hereby recommend as follows;
1. Radio stations, especially ABS, should perfect their *modus operandus* of news production and presentation with a view to changing society for the better.
2. Also, all the radio stations in Awka should note that most of the residents of the capital city tune in to their programs every day purposely to gather the information that will enrich their lives and transform the society at large. Thus, the proprietors, news reporters, editors, program directors, presenters, engineers on duty, announcers on duty, etc., should intensify effort towards meeting their audience demands academically and otherwise.
3. Similarly, it is of paramount importance for the National Broadcasting Commission (NBC) to double its efforts towards checkmating the activities of broadcast media in the country to ensuring their conformity to the ethics of the journalism profession. This will also help them to discover and sanction erring radio stations in order to bring sanity to the industry.
4. But more importantly, the society and listeners, students and teachers, schools and proprietors/government and private inclusive should henceforth start making judicious use of the media for academic and other purposes.